\RequirePackage{fix-cm}
\documentclass[twocolumn,final]{svjour3}          
\smartqed  
\usepackage{graphicx}
\usepackage{mathptmx}      
\usepackage{amsfonts, amssymb, amsmath}
\usepackage{hyperref}
\usepackage{pstool}
\usepackage[numbers,sort&compress]{natbib}

\newcommand{\changed}[1]{\textcolor{black}{#1}}
\newcommand{\md}{\text{d}}
\begin{document}

\title{1D model of precursors to frictional stick-slip motion allowing for robust comparison with experiments
}


\author{David Sk\aa lid Amundsen      \and
        Julien Scheibert              \and
        Kjetil Th\o gersen            \and
        J\o rgen Tr\o mborg           \and
        Anders Malthe-S\o renssen
}


\institute{D. Sk\aa lid Amundsen . J. Scheibert . K. Th\o gersen . J. Tr\o mborg . A. Malthe-S\o renssen \at
              Physics of Geological Processes, University of Oslo, P.O. Box 1048 Blindern, 0316 Oslo, Norway \\
             \emph{Present address:}\\
           J. Scheibert \at
              Laboratoire de Tribologie et Dynamique des Syst\`emes, CNRS, Ecole Centrale de Lyon, Ecully, France \\
              \email{julien.scheibert@ec-lyon.fr}
}

\date{Received: date / Accepted: date}

\maketitle

\begin{abstract}
We study the dynamic behaviour of 1D spring-block models of friction when the external loading is applied from a side, and not on all blocks like in the classical Burridge-Knopoff-like models. Such a change in the loading yields specific difficulties, both from numerical and physical viewpoints. To address some of these difficulties and clarify the precise role of a series of model parameters, we start with the minimalistic model by Maegawa \textit{et al.} (Tribol. Lett. \textbf{38} 313, 2010) which was proposed to reproduce their experiments about precursors to frictional sliding in the stick-slip regime. By successively adding (i) an internal viscosity, (ii) an interfacial stiffness and (iii) an initial tangential force distribution at the interface, we manage to (i) avoid the model's unphysical stress fluctuations, (ii) avoid its unphysical dependence on the spatial resolution and (iii) improve its agreement with the experimental results, respectively. Based on the behaviour of this improved 1D model, we develop an analytical prediction for the length of precursors as a function of the applied tangential load. We also discuss the relationship between the microscopic and macroscopic friction coefficients in the model.

\end{abstract}


\section{Introduction}
\label{sec:Introduction}

The dynamics of frictional interfaces are crucial to many situations in mechanical engineering~\cite{Urbakh-Klafter-Gourdon-Israelachvili-Nature-2004, Bhushan-Springer-2008}, geosciences~\cite{Scholz-CUP-2002} or biology~\cite{Scheibert-Leurent-Prevost-Debregeas-Science-2009, Prevost-Scheibert-Debregeas-CommunIntegrBiol-2009}. Today, after decades of studies, the science of contacts under time-invariant loading conditions, \textit{e.g.} static contacts or steady sliding contacts~\cite{Johnson-CUP-1985,Persson-Springer-2000,Baumberger-Caroli-AdvPhys-2006}, has reached a high level of advancement which, in many instances, enables quantitative reproduction of global~\cite{Baumberger-Caroli-AdvPhys-2006, Wandersman-Candelier-Debregeas-Prevost-PhysRevLett-2011} or local~\cite{Scheibert-Leurent-Prevost-Debregeas-Science-2009,Scheibert-Prevost-Debregeas-Katzav-AddaBedia-JMechPhysSolids-2009,Candelier-Debregeas-Prevost-Sensors-2011} measurements. In contrast, the dynamics of contacts under rapidly evolving loads or during fast unstable motion like stick-slip~\cite{Baumberger-Caroli-Ronsin-PhysRevLett-2002,Xia-Rosakis-Kanamori-Science-2004,Nielsen-Taddeucci-Vinciguerra-GeophysJInt-2010} is far less understood. In particular, recent experiments on the transition from static to kinetic friction of side-driven poly(methyl methacrylate) (PMMA) rough samples forming line contacts with a rough PMMA substrate have revealed unexpected features~\cite{Rubinstein-Cohen-Fineberg-Nature-2004, Rubinstein-Cohen-Fineberg-PhysRevLett-2007, Maegawa-Suzuki-Nakano-TribolLett-2010, BenDavid-Cohen-Fineberg-Science-2010}. The transition occurs through the fast (comparable to the speed of sound) propagation of micro-slip fronts through the contact~\cite{Rubinstein-Cohen-Fineberg-Nature-2004,BenDavid-Cohen-Fineberg-Science-2010}. It can also be preceded by a series of fronts that arrest before having ruptured the whole contact, thus denoted as precursors~\cite{Rubinstein-Cohen-Fineberg-PhysRevLett-2007, Maegawa-Suzuki-Nakano-TribolLett-2010}.

These results, which may have important implications for \textit{e.g.} the study of earthquakes, have triggered an active modelling activity. Braun \textit{et al.}~\cite{Braun-Barel-Urbakh-PhysRevLett-2009}, using a one-dimensional (1D) spring-block model with a complex time-dependent friction law, produced three types of micro-slip front velocities, analogous to that observed in~\cite{Rubinstein-Cohen-Fineberg-Nature-2004}. Maegawa \textit{et al.}~\cite{Maegawa-Suzuki-Nakano-TribolLett-2010}, using a 1D spring-block model with a simple Amontons-Coulomb (AC) friction law, showed that the length of precursors is modified when the external normal load is made asymmetric. Scheibert and Dysthe~\cite{Scheibert-Dysthe-EPL-2010}, using a quasi-static 1D model with AC friction, showed how the increasing tangential load itself induces an increasing pressure asymmetry which influences the precursors' series. Due to the intrinsic limitations of 1D models, all these studies yielded only a qualitative agreement with experiments. Very recently, Tr{\o}mborg \textit{et al.} demonstrated, using a 2D spring-block model with AC friction, that quantitative agreement with the kinematics (\textit{i.e.} the properties of the states in which no micro-slip front is propagating) of the experiments requires an accurate description of the interfacial stresses, and therefore the use of realistic boundary conditions on the sample~\cite{Tromborg-Scheibert-Amundsen-Thogersen-MaltheSorenssen-PhysRevLett-2011}.

Compared to such a 2D model, the strongest advantage of 1D models is that their results are much easier to analyse and understand, so that they provide opportunities for theoretical approaches. They also require simpler implementation and lower computational power. In most situations, in which very accurate results are not needed and/or a thorough qualitative understanding of the behaviour of the system is desired, 1D models are preferable. From the pioneering work of Burridge and Knopoff~\cite{Burridge-Knopoff-BSSA-1967}, \changed{spring-block models of friction have been extensively studied (see \textit{e.g.}~\cite{Carlson-Langer-PhysRevLett-1989, Olami-Feder-Christensen-PhysRevLett-1992, Braun-Peyrard-PhysRevLett-2008, Filippov-Popov-TribolInt-2010})}. These models have mainly been used to describe the statistical properties of the series of earthquakes at a seismic fault. Inertial blocks are connected in series via internal springs that model the crust's elasticity. A homogeneous tectonic loading is modelled by coupling, via loading springs, all individual blocks to the same rigid driving body. Such statistical analysis of homogeneously driven systems contrast with the recent 1D studies by Braun \textit{et al.}~\cite{Braun-Barel-Urbakh-PhysRevLett-2009} and Maegawa \textit{et al.}~\cite{Maegawa-Suzuki-Nakano-TribolLett-2010}, in which the time evolution of side-driven systems is analysed deterministically, in order to produce data comparable to the experimental measurements. The two models by Braun \textit{et al.} and Maegawa \textit{et al.} are actually very different. Both consider an array of blocks connected by internal springs, but the model by Braun \textit{et al.}~\cite{Braun-Barel-Urbakh-PhysRevLett-2009} also considers viscous dissipation and a complex time-dependent friction law emerging from the collective behaviour of interfacial springs with random stiffness, breaking threshold and reattachment time. In contrast the model by Maegawa \textit{et al.}~\cite{Maegawa-Suzuki-Nakano-TribolLett-2010}, which only considers blocks and springs and the minimalistic AC friction law, is probably the simplest possible model.

On the one hand, we will see that, due to its extreme simplicity, the model by Maegawa \textit{et al.}~\cite{Maegawa-Suzuki-Nakano-TribolLett-2010} yields results that are strongly resolution dependent, which prevents robust comparison with experiments. On the other hand, it is difficult to disentangle the respective roles of the many parameters of the model by Braun \textit{et al.}~\cite{Braun-Barel-Urbakh-PhysRevLett-2009}. The scope of this article is therefore to (i) construct, step by step, a minimal 1D side-driven spring-block model, the results of which are essentially resolution independent and (ii) qualify the capabilities of this model to reproduce the main qualitative features of recent experimental observations. 

This Letter is organised as follows. In section 2, we describe the model by Maegawa \textit{et al.}~\cite{Maegawa-Suzuki-Nakano-TribolLett-2010} and show its limitations. In section 3, we improve this model by introducing successively an internal viscosity, an interfacial stiffness and an initial tangential force distribution. In section 4, we describe an analytic prediction for the length of precursors in both Maegawa \textit{et al.}'s and our improved model.


\section{The model by Maegawa \textit{et al.} and its limitations}
\label{sec:TheMaegawaModelAndItsProblems}

\subsection{The model by Maegawa \textit{et al.}~\cite{Maegawa-Suzuki-Nakano-TribolLett-2010}}
\label{sec:TheMaegawaModel}

\begin{figure}[h]
\centering
\includegraphics{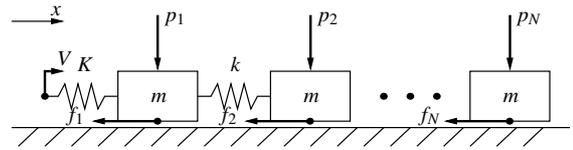}
\caption{Schematics of the model system. The sample is modeled by $N$ blocks of mass $m$ connected in series through springs of stiffness $k$. The trailing edge of the system (block 1) is slowly driven through a loading spring of stiffness $K$. Each block is also submitted to a normal force $p_i$ and to a friction force $f_i$.}
\label{fig:MaegawaModel}
\end{figure}

In the model developed by Maegawa~\textit{et al.}~\cite{Maegawa-Suzuki-Nakano-TribolLett-2010}, the slider is modelled as a chain of blocks connected by springs (Fig.~\ref{fig:MaegawaModel}), with material spring constant $k$ and block mass $m = M/N$, where $M$ is the total mass of the slider and $N$ is the number of blocks. In experiments, the base (also called track) is  fixed on a very stiff support, and it is therefore modelled as a rigid surface for simplicity. The tangential force is applied at the trailing edge of the system through a loading spring with stiffness $K$. One end of this spring is attached to the trailing edge block (block 1), while the other end of the spring moves at a constant velocity $V$. The normal force $p_i$ is imposed on each block, satisfying the criterion $\sum_{i=1}^N p_i = F_N$, where $F_N$ is the total applied normal force.

The equations of motion are given by
\begin{equation}
m \ddot{u}_n = \left\{ \begin{array}{ll}
k(u_2 - u_1) + F_T + f_1, & n = 1 \\
k(u_{n+1} - 2u_n + u_{n-1}) + f_n, & 2 \leq n \leq N-1 \\
k(u_{N-1} - u_N) + f_N, & n = N,
\end{array} \right.
\end{equation}
where $u_n = u_n(t)$ is the position of block $n$ as a function of time relative to its equilibrium position and $\ddot{}$ denotes the double derivative with respect to $t$. $F_T = F_T(t)$ is the driving force (or tangential force/load) given by
\begin{equation}
F_T = K(Vt - u_1) .
\end{equation}

A local friction law giving the friction forces $f_n$ is imposed between the blocks and the base. AC friction is used with local kinetic and static friction coefficients $\mu_k$ and $\mu_s$, respectively. The resulting global friction coefficients are similarly denoted by $\mu_K$ and $\mu_S$. The friction force on block $n$, $f_n$, is then given by
\begin{equation}
f_n = \left\{ \begin{array}{ll}
\leq \mu_s p_n, & \dot{u}_n = 0 \\
-\text{sgn}(\dot{u}_n) \mu_k p_n, & \dot{u}_n \neq 0 ,
\end{array} \right.
\end{equation}
where, when $\dot{u}_n = 0$, equilibrium of block $n$ imposes that $f_n$ balances all other forces acting on block $n$.

The material spring constant $k$ is chosen such that the elastic deformation of the model is similar to that of a linear elastic medium with Young's modulus $E$, which yields:
\begin{equation}
k = (N-1)E S /L ,
\label{eq:k}
\end{equation}
where $S$ and $L$ are the cross-section area and the length of the slider, respectively.

In the experiments by Maegawa \textit{et al.}~\cite{Maegawa-Suzuki-Nakano-TribolLett-2010}, an asymmetric normal loading was used, leading to the following linear model for $p_n$:
\begin{equation}
p_n = \frac{F_N}{N} \left( 1 - \frac{2n - N - 1}{N - 1} \theta \right),
\label{eq:p_n}
\end{equation}
where $\theta \in [-1,1]$ is a measure of the non-uniformity in the normal loading.

The values of the parameters are chosen to be in agreement with their respective values in the experiments by Maegawa \textit{et al.}~\cite{Maegawa-Suzuki-Nakano-TribolLett-2010}, which are: $K = 0.8$~MN/m, $V = 0.1$~mm/s, $F_N = 400$~N, $M = 0.012$~kg, $L = 100$~mm, $S=100$~mm$^2$, $E = 2.5$~GPa, $\mu_s = 0.7$ and $\mu_k = 0.45$.


\subsection{Results and limitations of the model by Maegawa \textit{et al.}}
\label{sec:ResultsAndProblemsOfTheMaegawaModel}

\begin{figure}
\centering
\includegraphics{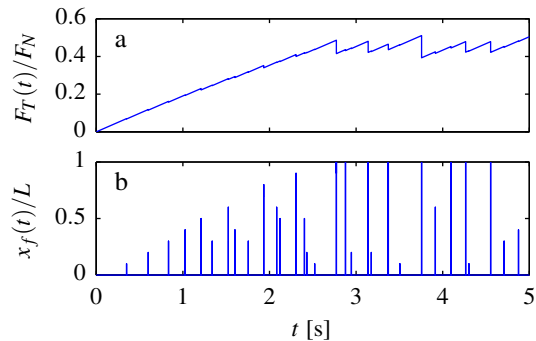}
\caption{Time evolution of $F_T$ (a) and $x_f$ (b) in the model by Maegawa \textit{et al.} using $N = 10$ (as in~\cite{Maegawa-Suzuki-Nakano-TribolLett-2010}) and $\theta = 0$. Before macroscopic stick-slip (reached when $t \approx$ 2.8 s), the loading curve is punctuated by partial relaxations associated with precursors to sliding, \textit{i.e.} micro-slip fronts spanning a length smaller than $L$.}
\label{fig:loadingCurveLengthPrecursorsAmontonsCoulombN=10}
\end{figure}

We have implemented the model by Maegawa \textit{et al.} and tested our code by comparing it to (i) two other similar codes and (ii) the analytical solution of the equation of motion for a one-block system. The simulation starts with each spring at its equilibrium length, \textit{i.e.} $u_n = 0$ for all blocks. From $t = 0$, the force from the driving spring on block $1$ is increased. Figure~\ref{fig:loadingCurveLengthPrecursorsAmontonsCoulombN=10}a shows that the resulting time evolution of $F_T (t)$ exhibits stick-slip behaviour. Each time a drop in the driving force is observed, some part of the slider moves relative to the base. We call the short time intervals during which movement occurs \textit{events}.

Between events, no block is moving. Since the driving force is applied only at the trailing edge block, only block 1 is loaded and eventually reaches its static friction threshold, so that all events must nucleate at the trailing edge. The movement of block 1 then loads block 2, which itself reaches its threshold and so on. This succession of blocks starting to move defines a micro-slip front, which propagates towards the leading edge, in analogy with the fronts observed in experiments~\cite{Rubinstein-Cohen-Fineberg-Nature-2004, Maegawa-Suzuki-Nakano-TribolLett-2010}. The distance from the trailing edge to the micro-slip front as a function of time, $x_f(t)$, is shown in Fig.~\ref{fig:loadingCurveLengthPrecursorsAmontonsCoulombN=10}b. If this front reaches the leading edge, the event is a global event, and the whole slider moves relative to the base.

From Fig.~\ref{fig:loadingCurveLengthPrecursorsAmontonsCoulombN=10} it is evident that not all events are global: smaller events are observed between global events. In addition, a series of events with increasing maxima of $x_f(t)$ is seen to precede the first global event. These events occur for $F_T$ well below the macroscopic static friction threshold, and are called \textit{precursors}~\cite{Rubinstein-Cohen-Fineberg-PhysRevLett-2007,Maegawa-Suzuki-Nakano-TribolLett-2010}. The maximum of $x_f(t)$ during a precursor event, \textit{i.e.} the length of a precursor, is denoted by $L_p$.

To perform a quantitative comparison between their experimental and numerical results, Maegawa \textit{et al.} focused on the relationship between the normalized length $L_p/L$ of the series of precursors (Fig.~\ref{fig:loadingCurveLengthPrecursorsAmontonsCoulombN=10}b) and the normalized tangential force $F_T/F_N$ at which they are triggered (Fig.~\ref{fig:loadingCurveLengthPrecursorsAmontonsCoulombN=10}a). To do this, they discarded all simulated precursors having a length smaller than any of the previous events, with the justification that no such smaller event was observed in the experiments. Note that 2D models produce series of precursors of monotonically increasing length, so that practically no event has to be discarded~\cite{Tromborg-Scheibert-Amundsen-Thogersen-MaltheSorenssen-PhysRevLett-2011}.

\begin{figure}
\centering
\includegraphics{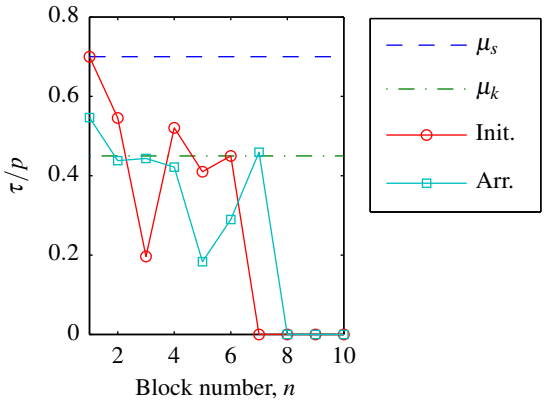}
\caption{Tangential force distribution at initiation of a microslip-event ($\circ$). Block 1 has reached its slip threshold ($\tau / p = \mu_s$). The event involves all blocks from 1 to 6 (\textit{i.e.} a precursor of length $L_p/L = 0.6$), leading to a modified tangential force distribution at arrest ($\square$). Results obtained using $N = 10$ (as in~\cite{Maegawa-Suzuki-Nakano-TribolLett-2010}) and $\theta = 0$.}
\label{fig:shearForceProfileNoVistosityAmontonsCoulombN=10}
\end{figure}

Figure~\ref{fig:shearForceProfileNoVistosityAmontonsCoulombN=10} shows the tangential force $\tau$ normalised by the normal force $p$ on each block at the time of initiation and arrest of a precursor of length $L_p = 0.6L$. The tangential force is here defined as the total force on a block excepting the friction force. Each event is initiated when the tangential force on block $1$ reaches the local static friction threshold. As block $1$ moves, the tangential force on block $2$ increases, eventually reaching the local static friction threshold, and starts to move. The precursor event arrests when the tangential force built on a block by its left neighbour is not sufficient to make it reach its static friction threshold. The slow loading of block $1$ continues, and will eventually trigger a new event nucleating at the trailing edge.

\begin{figure}
\centering
\includegraphics{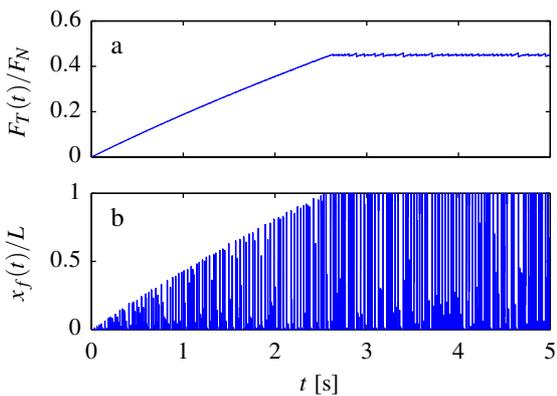}
\caption{Time evolution of $F_T$ (a) and $x_f$ (b) in the model by Maegawa \textit{et al.} using $N = 100$ and $\theta = 0$. Comparison with Fig.~\ref{fig:loadingCurveLengthPrecursorsAmontonsCoulombN=10} shows a drastic increase in the time-frequency of micro-slip events.}
\label{fig:loadingCurveLengthPrecursorsAmontonsCoulombN=100}
\end{figure}

The spatial resolution used in the above results, \textit{i.e.} $N = 10$ as used by Maegawa \textit{et al.}~\cite{Maegawa-Suzuki-Nakano-TribolLett-2010}, is rather low. This is especially evident in Fig.~\ref{fig:shearForceProfileNoVistosityAmontonsCoulombN=10}, which only contains $10$ data points for $\tau/p$. We expected to improve these results by simply increasing the resolution to $N = 100$. The corresponding loading and front position curves, $F_T(t)$ and $x_f(t)$, are shown in Fig.~\ref{fig:loadingCurveLengthPrecursorsAmontonsCoulombN=100}.

One of the limitations of the model by Maegawa \textit{et al.} is now clearly evident. By only changing the spatial resolution, the loading and front position curves are changed significantly. Some aspects are unchanged: precursors precede the first global event and then stick-slip behaviour is observed. The final average level of $F_T/F_N$ also appears to be conserved. However, the amplitude of the drops in the loading curve is reduced, while the number of events, both global and precursory, is seen to increase significantly.

\begin{figure}
\centering
\includegraphics{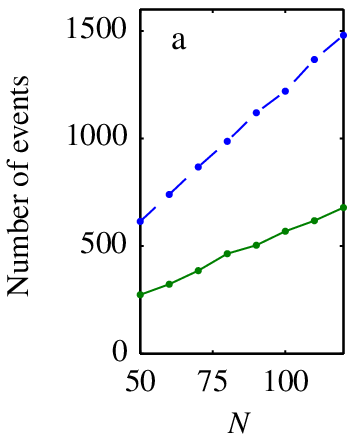}
\includegraphics{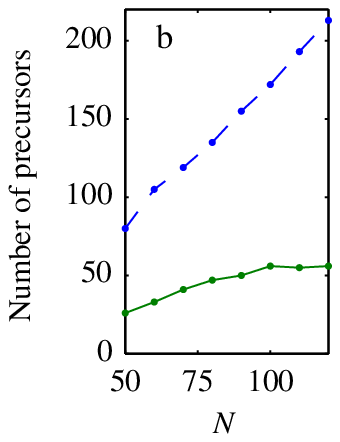}
\caption{Number of different kinds of events as a function of $N$, using $\theta = 0$. (a) Number of global events (solid line) for $t \in [5\text{ s}, 20\text{ s}]$ and total number of events (dashed line). (b) Number of precursors longer than any previous one (solid line) and total number of precursors (dashed line). All curves are stongly increasing functions of $N$, indicating unphysical resolution-dependence of the model's results.}
\label{fig:numberOfEventsAmontonsCoulomb}
\end{figure}

To illustrate this scaling with respect to the model resolution, we have plotted in Fig.~\ref{fig:numberOfEventsAmontonsCoulomb} the evolution with $N$ of (i) the total number of events and the number of global events (Fig.~\ref{fig:numberOfEventsAmontonsCoulomb}a), and (ii) the total number of precursors and the number of precursors longer than any previous ones (Fig.~\ref{fig:numberOfEventsAmontonsCoulomb}b). An approximately linear increasing trend is observed in all four curves. This behaviour is problematic as soon as one wants to compare to experiments, in which the size of the drops in $F_T/F_N$ and the number of events are well-defined experimental measurements. Maegawa \textit{et al.}~\cite{Maegawa-Suzuki-Nakano-TribolLett-2010} used $N=10$, which produced a number of precursors similar to that observed in their experiments, but this agreement appears to be casual. A robust model should produce almost identical numbers of events whatever the spatial discretization of the slider. In this respect, note that 2D models do produce a resolution-independent number of events~\cite{Tromborg-Scheibert-Amundsen-Thogersen-MaltheSorenssen-PhysRevLett-2011}.

\begin{figure}
\centering
\includegraphics{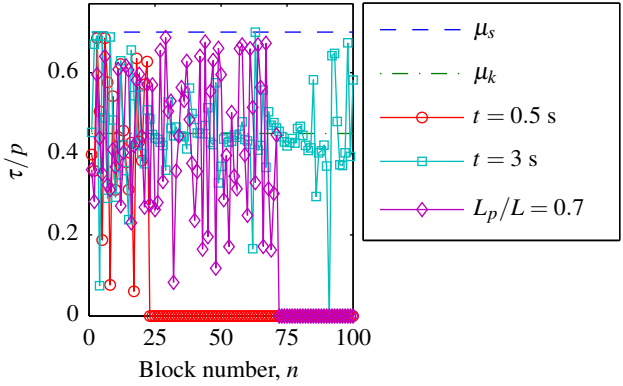}
\caption{Tangential force distribution $\tau_n$ normalised by the normal force $p_n$ on each block, at three different times: at $t = 0.5$ s, $t = 3$ s, and at the arrest of the precursor of length $L_p/L = 0.7$ using $N = 100$ and $\theta = 0$. Unwanted large oscillations with a half-period equal to the lattice spacing appear.}
\label{fig:shearForceProfileNoVistosityAmontonsCoulombN=100}
\end{figure}

Another problematic behaviour of the model affects the tangential force spatial distribution, as shown in Fig.~\ref{fig:shearForceProfileNoVistosityAmontonsCoulombN=100}. The tangential force has been plotted at three different times: at $t = 0.5$ s, $t = 3$ s, and at the arrest of the precursor of length $L_p/L = 0.7$. Strong oscillations are observed, with a half spatial period of the order of the lattice spacing, whatever the number of blocks $N$ used. Again this unphysical dependence on the model resolution impedes a robust comparison with experimental measurements of the tangential stress distribution $\tau(x)$, like \textit{e.g.} those of~\cite{BenDavid-Cohen-Fineberg-Science-2010}.

In their quantitative comparison between model and experimental results, Maegawa \textit{et al.}~\cite{Maegawa-Suzuki-Nakano-TribolLett-2010} focused on the length of precursors $L_p$ as a function of the tangential load $F_T$, $L_p(F_T)$. This relationship has also been studied experimentally by Rubinstein \textit{et al.}~\cite{Rubinstein-Cohen-Fineberg-PhysRevLett-2007}. While similar behaviours are found, we will only compare our model results to the experimental results of Maegawa \textit{et al.}.

The experimental setup by Maegawa \textit{et al.} allowed for non-uniformities in the normal loading, and they studied its consequences on the length of precursors. The non-uniformity in the normal loading is modelled as an asymmetric distribution of the normal loads $p_n$ by using Eq.~\eqref{eq:p_n}. The value of the parameter $\theta = \pm 0.833$ is chosen to be in agreement with its corresponding experimental value.

\begin{figure}
\centering
\includegraphics{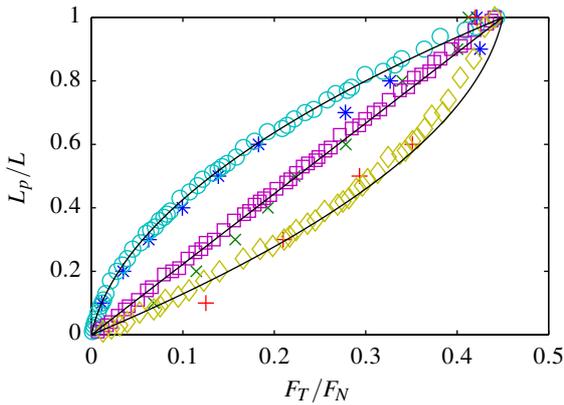}
\caption{Length of precursors $L_p$ normalised by the system length $L$ as a function of the tangential load $F_T$ at event arrest normalised by the normal load $F_N$ for $N = 10$ and $\theta = 0.833$ ($\bigcirc$), $\theta = 0$ ($\square$) and $\theta = -0.833$ ($\Diamond$), and $N = 100$ and $\theta = 0.833$ ($*$), $\theta = 0$ ($\times$) and $\theta = -0.833$ ($+$). Solid lines are the analytical predictions described in Sec.~\ref{sec:AnalyticPredictionOfPrecursorLengths}.}
\label{fig:precursorLengthsNoInitialShearForceAmontonsCoulomb}
\end{figure}

The length of precursors $L_p$ is plotted as a function of the tangential force $F_T$ at event arrest in Fig.~\ref{fig:precursorLengthsNoInitialShearForceAmontonsCoulomb}. Three different values of $\theta$ are used, and the results for both $N = 10$ and $N = 100$ are included. As discussed above, the number of precursors increases with $N$. However, the shape of the curves does not change significantly with $N$, thus enabling comparison with the shape of the experimental curves.

The qualitative behaviour of $L_p$ as a function of $F_T$ is in agreement with experiments. For $\theta = 0.833$, the normal force on the trailing edge is reduced, leading to longer precursors for the same tangential load compared to $\theta = 0$. On the other hand, $\theta = -0.833$ leads to an increased normal load on the trailing edge, and precursors are shorter for a given $F_T$. All curves converge to the same point at $L_p/L = 1$, meaning that the \textit{global} static friction threshold $\mu_S$ (the value of $F_T / F_N$ when the first global event occurs) is independent of the normal loading \changed{distribution}. The value of $\mu_S$ appears to be approximately $0.45$, which incidentally is the value of the local kinetic friction coefficient $\mu_k$. The reason for this will become clear in Sec.~\ref{sec:AnalyticPredictionOfPrecursorLengths}. A quantitative comparison between the experimental results in~\cite{Maegawa-Suzuki-Nakano-TribolLett-2010} and the model results in Fig.~\ref{fig:precursorLengthsNoInitialShearForceAmontonsCoulomb}, however, reveals large discrepancies: all three experimental curves are found way below their simulated counterpart, meaning that the simulation strongly overestimates $L_p$ for any given $F_T$; the rapid increase in precursor length after $L_p/L \approx 0.5$ that is observed experimentally has no equivalent in the model.

Summing up, three main limitations of the model by Maegawa \textit{et al.} have been identified: (i) the tangential force shows large oscillations, the wavelength of which is controlled by the lattice spacing, (ii) the number of all kinds of events is an increasing function of $N$ and (iii) the quantitative agreement with the $L_p(F_T)$-curve between model and experiments is poor. In the following section, we propose improvements of the model that contribute to overcome these limitations.


\section{Improvements of the model by Maegawa \textit{et al.}}


\subsection{Introducing a relative viscous damping}

Resolution dependent oscillations are known to occur in Burridge-Knopoff-like models and more generally in dynamic rupture models involving AC friction at the interface between dissimilar elastic media~\cite{Adams-JApplMech-1995}. Classical ways to reduce them significantly are either to regularize the AC friction law (see \textit{e.g.} \cite{BenZion-JMechPhysSolids-2001} and references therein) or to introduce a viscous damping in the system~\cite{Myers-Langer-PRE-1993, Shaw-GRL-1994, Knopoff-Ni-GJI-2001, Braun-Barel-Urbakh-PhysRevLett-2009, Tromborg-Scheibert-Amundsen-Thogersen-MaltheSorenssen-PhysRevLett-2011}. \changed{Both are physically sound, but we choose to} adopt the second approach. Physically, such viscosity is a way to model the energy dissipation that any material undergoes during deformation. After Knopoff and Ni~\cite{Knopoff-Ni-GJI-2001}, we consider the following form for the viscous force $F_n^\eta$:
\begin{equation}
F_n^\eta = \left\{ \begin{array}{ll}
\eta \left( \dot{u}_{2} - \dot{u}_1 \right), & n = 1 \\
\eta \left( \dot{u}_{n+1} - 2\dot{u}_n + \dot{u}_{n-1} \right), & 2 \leq n \leq N-1 \\
\eta \left( \dot{u}_{N-1} - \dot{u}_N \right), & n = N ,
\end{array} \right.
\label{eq:F_n^eta}
\end{equation}
which is a damping on the relative motion of neighbouring blocks. \changed{As in \textit{e.g.}~\cite{Maegawa-Suzuki-Nakano-TribolLett-2010, Tromborg-Scheibert-Amundsen-Thogersen-MaltheSorenssen-PhysRevLett-2011}, we assume that energy dissipation due to the motion of a block relative to the substrate is satisfactorily included in $f_n$, and therefore do not, in contrast to \textit{e.g.}~\cite{Braun-Barel-Urbakh-PhysRevLett-2009, Braun-Tosatti-PhilosMag-2011}, include any viscous damping at the interface in our system. This also serves to keep the model as simple as possible.} The equations of motion are then given by
\begin{equation}
m \ddot{u}_n = \left\{ \begin{array}{ll}
k(u_2 - u_1) + F_T + F_1^\eta + f_1, & n = 1 \\
k(u_{n+1} - 2u_n + u_{n-1}) + F_n^\eta + f_n, & 2 \leq n \leq N-1 \\
k(u_{N-1} - u_N) + F_N^\eta + f_N, & n = N.
\end{array} \right.
\label{eq:eqOfMotionWithViscosity}
\end{equation}

The tangential force $\tau_n$ is still defined as the sum of all forces on a block excepting the friction force, \textit{i.e.} now including $F_n^\eta$. It remains to choose a reasonable value of the damping coefficient $\eta$.

In App.~\ref{sec:RelativeViscousDampingInALinearChainOfBlocks}, the value of $\eta$ for critical damping of waves of wavelength $\lambda = 2a$, corresponding to the cut-off wavelength, is calculated. The result is $\eta_c = \sqrt{km}$. It is also shown that the values of $\eta$ corresponding to critical damping of higher wavelength oscillations are always larger than $\sqrt{km}$. We want to damp out oscillations of wavelength $\lambda = 2a$, and using $\eta = \sqrt{km}$ is then a possibility. However, waves of other wavelengths close to $\lambda = 2a$ will also be highly damped, causing significant changes to the dynamics. Since this is an undesirable effect, a compromise has to be made. As suggested in~\cite{Knopoff-Ni-GJI-2001}, the value
\begin{equation}
\eta = \sqrt{0.1}\sqrt{km} \approx 0.32\sqrt{km}
\label{eq:eta}
\end{equation}
is used in the following. Note that since $k \propto N-1$ and $m \propto 1/N$, $\eta$ is $N$-independent for $N \gg 1$.

\begin{figure}
\centering
\includegraphics{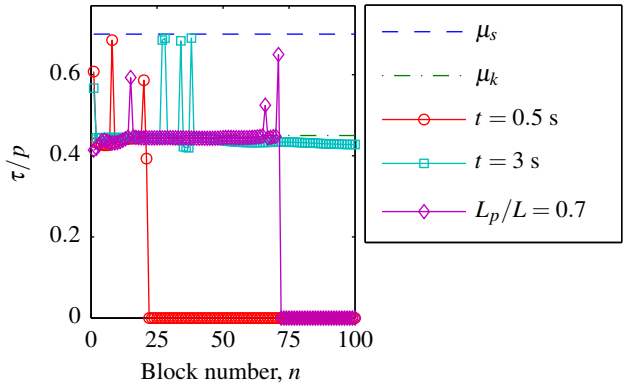}
\caption{The tangential force $\tau$ normalised by the normal force $p$ on each block is plotted at three different times: at $t = 0.5$ s, $t = 3$ s, and at the arrest of the precursor of length $L_p/L = 0.7$ from a simulation including the relative viscous damping and using $N = 100$, $\theta = 0$ and $\eta = \sqrt{0.1}\sqrt{km}$. Comparison with Fig.~\ref{fig:shearForceProfileNoVistosityAmontonsCoulombN=100} shows that lattice-controlled fluctuations have disappeared.}
\label{fig:shearForceProfileWithVistosityAmontonsCoulombN=100}
\end{figure}

Figure \ref{fig:shearForceProfileWithVistosityAmontonsCoulombN=100} shows the tangential force distribution obtained when relative viscous damping is included. The improvement with respect to Fig.~\ref{fig:shearForceProfileNoVistosityAmontonsCoulombN=100} is clear: the short wavelength oscillations have almost disappeared, resulting in a physically reasonable smooth tangential force profile. $\tau/p$ also appears to be on average approximately equal to $\mu_k$ in ruptured regions, a fact that will be utilised below to predict the precursor lengths.

Some small one- or two-node spikes remain. They have a different origin, as they are mainly caused by the discreteness of the friction law: at the tip of the rupture, one block is moving and therefore increasing the force on its neighbour, which is still stationary. The region beyond this stationary block is therefore not affected by the approaching rupture front, and a spike will therefore appear at the rupture front. Spikes may also appear as an event arrests, also caused by one part of the interface slipping while another part is stuck.

In all the following, the viscous damping introduced in this section will be used.


\subsection{Introducing a tangential stiffness of the interface}
\label{sec:TheSpringToTrackStaticFrictionLaw}

As discussed above, the model by Maegawa \textit{et al.} exhibits an unphysical scaling with $N$. It is possible to understand this scaling by considering how the system is tangentially loaded. As stated above, the driving force only acts on block $1$. In order for an event to nucleate, this block has to reach its static friction threshold, which is proportional to the normal force: $\mu_s p_1 \propto p_1 \propto 1/N$. Since the added driving force per time is independent of $N$, the time between two events will be proportional to $1/N$, and the frequency of events consequently scales as $N$, an argument which is fully consistent with Fig.~\ref{fig:numberOfEventsAmontonsCoulomb}. The origin of the odd $N$-\changed{dependence} of the model by Maegawa \textit{et al.} is therefore the decreasing size of the loading region as $N$ is increased.

In a physical system, like in the experiments by Maegawa \textit{et al.}, the loading region has a well defined spatial extension, which is a combination of various effects. First, the tangential loading is applied at the trailing edge of the slider at some effective height $h$ above the interface. As discussed experimentally in~\cite{Rubinstein-Cohen-Fineberg-PhysRevLett-2007} and modelled in~\cite{Tromborg-Scheibert-Amundsen-Thogersen-MaltheSorenssen-PhysRevLett-2011}, such a loading condition makes the tangential stress very high in the region near the trailing edge, the extension of which is of order $h$. Second, the interface between slider and base is not rigid. Both surfaces are rough and the multi-contact layer between them has a finite tangential stiffness associated to the tangential deformation of the microasperities involved in the contact. Such stiffness can be measured experimentally~\cite{Berthoud-Baumberger-ProcRSocA-1998} and is found much smaller than the slider's bulk stiffness. Such a low interfacial stiffness can be responsible for deviations with respect to AC friction in static contacts~\cite{Scheibert-Prevost-Frelat-Rey-Debregeas-EPL-2008}. The interfacial stiffness also results in a physical finite size of the loading region, since a localized tangential force at the interface will induce tangential strains within the rough layer not only at the loaded point but also in its neighbourhood. 

\begin{figure}
\centering
\includegraphics{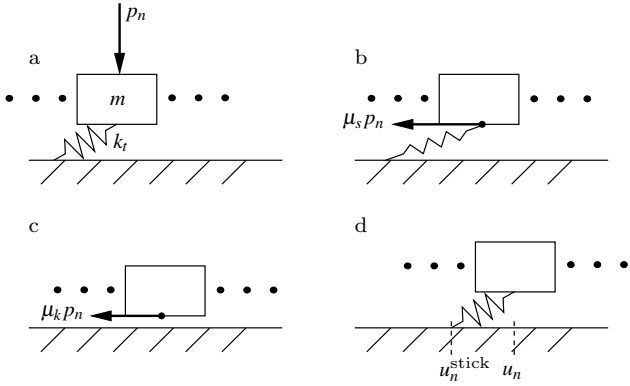}
\caption{Sketch of the behaviour of the spring between block $n$ and the track. (a) The static friction force on block $n$ is exerted through a spring of stiffness $k_t$ attached to the track. (b) As the block is moving, the spring is stretched until it reaches its breaking strength $\mu_s p_n$. (c) When the spring is broken, block $n$ is subject to the kinetic friction force $\mu_k p_n$. (d) As the block stops, the spring reattaches at $x=u_n^{stick}$ such that the total force on the block is zero at the time of reattachment.}
\label{fig:springToTrackEvolution}
\end{figure}

The first effect relates to the elastic coupling of points of the interface through the slider's bulk, which cannot be accounted for explicitly by a 1D model like the one considered here. In contrast, the effect of the interfacial stiffness can be introduced\changed{~\cite{Braun-Tosatti-PhilosMag-2011}} in a 1D model in the following way. Each block is initially attached to the track by a spring with stiffness $k_t$ as seen in Fig.~\ref{fig:springToTrackEvolution}a. The spring connecting block $n$ to the track has a breaking strength $\mu_s p_n$ (Fig.~\ref{fig:springToTrackEvolution}b). When the spring to the track is detached the block is subject to the kinetic friction force $\pm \mu_k p_n$ (Fig.~\ref{fig:springToTrackEvolution}c). As the velocity of this block reaches zero, the spring reattaches such that the total force on the block is $0$ at the time of reattachment (Fig.~\ref{fig:springToTrackEvolution}d).

The friction force is now given by
\begin{equation}
f_n = \left\{ \begin{array}{ll}
-k_t \left( u_n - u_n^\text{stick} \right) & \text{if attached}, \\
-\text{sign}\left( \dot{u}_n \right)\mu_k p_n & \text{if detached} ,
\end{array} \right.
\end{equation}
where $u_n^\text{stick}$ is the attachment point of the spring between block $n$ and the track. It is given by
\begin{equation}
u_n^\text{stick} = u_n - \frac{\tau_n}{k_t},
\label{eq:u_n^stick}
\end{equation}
where $u_n$ and $\tau_n$ are the position of and tangential force on block $n$ at the instant of its last reattachment to the track. This causes the total force on block $n$ to be zero at the time of reattachment. The spring then detaches at the time step in which one finds
\begin{equation}
\left|-k_t \left( u_n - u_n^\text{stick} \right)\right| > \mu_s p_n .
\end{equation}
The equations of motion are still given by Eqs.~\eqref{eq:eqOfMotionWithViscosity}.

\begin{figure}
\centering
\includegraphics{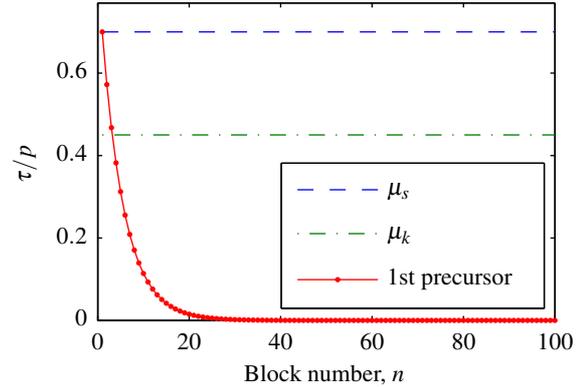}
\caption{Tangential force $\tau$ normalised by the normal force $p$ at initiation of the first precursor, when an interfacial stiffness is considered. The first block has just reached it threshold for slip ($\tau / p= \mu_s$). The tangential force decays exponentially according to Eq.~\eqref{eq:shearForceProfile}. Results obtained using $N = 100$, $\theta = 0$ and $k_t = 10^{7}$~N/m.}
\label{fig:shearForceProfileFirstPrecursorSpringToTrack}
\end{figure}

As the system is loaded tangentially, a finite region around the driving point is affected. This is illustrated in Fig.~\ref{fig:shearForceProfileFirstPrecursorSpringToTrack} where the tangential force at the time of nucleation of the first precursor is plotted. The length of this region depends on the stiffness $k_t$ of the springs between the blocks and the track.

\changed{Assuming (i) $N \gg 1$, (ii) the length of the loading region to be much smaller than $L$ and (iii) slow loading compared to the internal dynamics of the system, we calculate analytically in App.~\ref{sec:ScalingOfTheModelWithTheSpringToTrackStaticFrictionLaw} the tangential force profile at the time of nucleation of an event. No assumption is made for the tangential force profile at the time of arrest of the previous event, and the calculation is therefore valid for all events, not only for the first precursor shown in Fig.~\ref{fig:shearForceProfileFirstPrecursorSpringToTrack}. The result is an exponential decay of the tangential force with a characteristic length $l_0$, which consequently is a measure of the size of the loading region. From App.~\ref{sec:ScalingOfTheModelWithTheSpringToTrackStaticFrictionLaw}, $l_0$ is related to $k_t$ by}
\begin{equation}
l_0 = \sqrt{\frac{E S L}{N k_t}}.
\label{eq:l0N}
\end{equation}
In this expression, we recognize $N k_t$ to be the total stiffness, $k^\text{tot}_t$, of the interface ($N$ springs of individual stiffness $k_t$ in parallel), which is a measurable quantity in a given experimental setup. We then obtain the relation 
\begin{equation}
l_0 = \sqrt{E S L /k_t^\text{tot}},
\end{equation}
which shows that the size $l_0$ of the loading region is now independent of $N$, \textit{i.e.} of the model's spatial resolution.

In the model by Maegawa \textit{et al.}, the simple rigid-plastic-like AC friction ruled the behaviour of the interface. Now that the interfacial stiffness is introduced, the friction law is elasto-plastic-like, an improvement that has often been defended as a necessary extension of AC~\cite{Berthoud-Baumberger-ProcRSocA-1998, Brzoza-Pauk-ArchApplMech-2008, Scheibert-Prevost-Frelat-Rey-Debregeas-EPL-2008, Nakano-Maegawa-TribolInt-2009, Braun-Barel-Urbakh-PhysRevLett-2009}. Our model is the simplest improvement of the model by Maegawa \textit{et al.} that accounts for interfacial stiffness.

\begin{figure}
\centering
\includegraphics{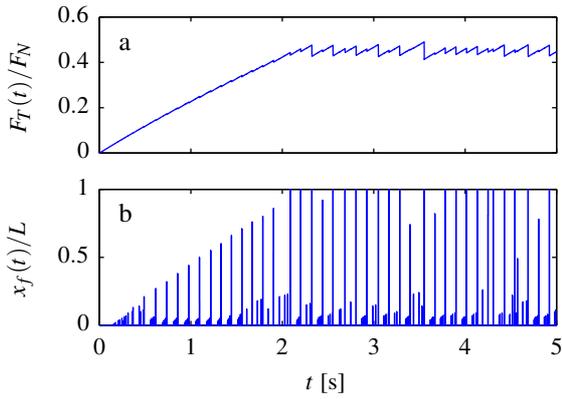}
\caption{\changed{Time evolution of $F_T$ (a) and $x_f$ (b) in our improved model including both a relative viscous damping and an interfacial stiffness, using $N = 100$ and $\theta = 0$. Comparison with Figs.~\ref{fig:loadingCurveLengthPrecursorsAmontonsCoulombN=100} and~\ref{fig:loadingCurveLengthPrecursorsAmontonsCoulombN=10} shows similar envelopes but very different numbers of events and amplitudes of the drops in $F_T/F_N$.}}
\label{fig:loadingCurveLengthPrecursorsTangentialStiffnessN=100}
\end{figure}

\changed{The envelopes of the time dependence of both the tangential load and length of precursors in this improved model are not changed significantly compared to those obtained using AC friction, as seen by comparing Figs.~\ref{fig:loadingCurveLengthPrecursorsAmontonsCoulombN=10} and~\ref{fig:loadingCurveLengthPrecursorsTangentialStiffnessN=100}. However, comparing Figs.~\ref{fig:loadingCurveLengthPrecursorsTangentialStiffnessN=100} to~\ref{fig:loadingCurveLengthPrecursorsAmontonsCoulombN=100} shows a significant decrease in the number of events and a consequent increase in the amplitude of the drops in $F_T/F_N$.} The shear force profiles are also very similar to that of Fig.~\ref{fig:shearForceProfileWithVistosityAmontonsCoulombN=100}, except that spikes now decay exponentially with a characteristic length $l_0$. We will now check the model's behaviour with respect to its scaling with $N$. To do this, one has to choose the interfacial stiffness $k_t$ (or equivalently a loading zone size $l_0$). The value of $k_t$ could be calibrated using an experimental measurement of the total interfacial stiffness $k_t^\text{tot}$. Here, we do not have access to such a measurement, so we rather exploit the fact that the number of precursors longer than any previous one is controlled by \changed{the parameter} $l_0$. By trial and error, we found that $l_0 = 5$~mm, corresponding to $k_t = 10^{7}$~N/m produced a number of such events similar to that observed in the experiments by Maegawa \textit{et al.}. We will use this value of $l_0$ in the following.

\begin{figure}
\centering
\includegraphics{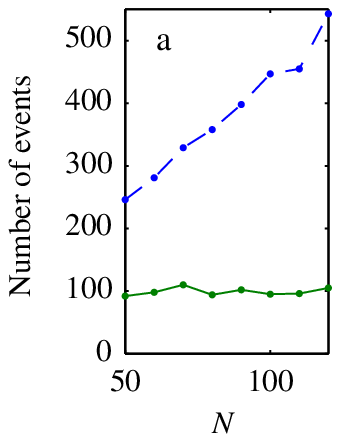}
\includegraphics{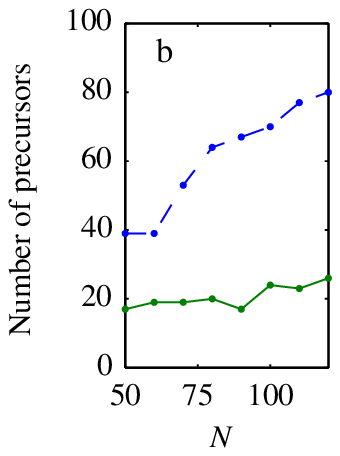}
\caption{Number of different kinds of events as a function of $N$, when the interfacial stiffness in taken into account. $\theta = 0$, $\eta = \sqrt{0.1} \sqrt{km}$ and $l_0 = 5$~mm. (a) Number of global events for $t \in [5\text{ s}, 20\text{ s}]$ (solid line) and total number of events (dashed line). (b) Number of precursors longer than any previous one (solid line) and total number of precursors (dashed line). Comparison with Fig.~\ref{fig:numberOfEventsAmontonsCoulomb} shows that introduction of an interfacial tangential stiffness suppresses the resolution dependence of the numbers of global events and of precursors longer than any previous one.}
\label{fig:numberOfEventsSpringToTrack}
\end{figure}

In Fig.~\ref{fig:numberOfEventsSpringToTrack} the number of events using the elasto-plastic friction law is plotted in the same way as in Fig.~\ref{fig:numberOfEventsAmontonsCoulomb} for AC friction. Both the number of global events and the number of precursors longer than any previous one are now seen to be approximately constant. The total number of events and the total number of precursors are, however, still increasing with $N$. This is mainly due to events involving one block, \textit{i.e.} events in which no real front propagation occurs. In other words, interfacial friction is found to satisfactorily solve the resolution dependence of the model's results, provided one considers events that have a measurable length.

\subsection{Introduction of an initial tangential force distribution}

\begin{figure}
\centering
\includegraphics{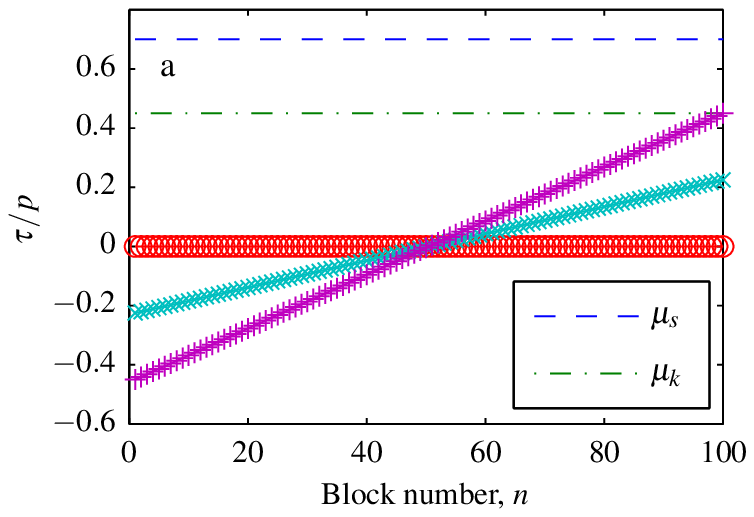} \\
\includegraphics{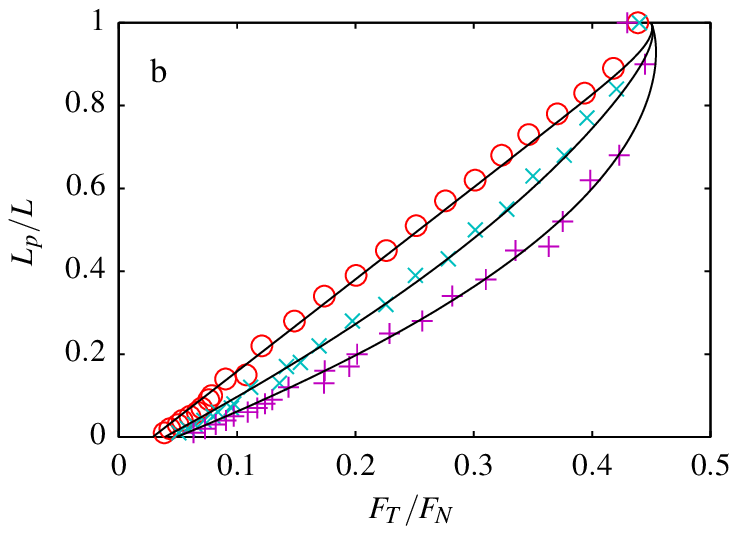}
\caption{(a) Three different antisymmetric initial tangential force distributions aimed at modelling the effect of friction-frustrated differential Poisson expansion during normal loading. $\circ$: homogeneous distribution. $\times$ and $+$: linear profiles of increasing slope ($\beta=0.225$ and 0.45 in Eq.~\eqref{eq:tau^0(x)} respectively). (b) The length of precursors corresponding to the three initial tangential force profiles shown in (a). Results obtained using $N = 100$, $\theta = 0$, $\eta = \sqrt{0.1}\sqrt{km}$ and $l_0 = 5$~mm. Solid lines are the analytical predictions of the precursor lengths discussed in Sec.~\ref{sec:AnalyticPredictionOfPrecursorLengths}.}
\label{fig:initialShearForce}
\end{figure}

The introduction of an internal viscosity and an interfacial stiffness in the model by Maegawa \textit{et al.} allowed us to obtain force distributons and numbers of micro-slip events that were physically sound. However, the predicted $L_p$ \textit{vs.} $F_T$ curves still follow the same shapes as those shown in Fig.~\ref{fig:precursorLengthsNoInitialShearForceAmontonsCoulomb} which, as already mentioned, deviate significantly from those obtained experimentally. In an effort to further improve the 1D model, we note that one of the main differences between the model and the experiment is the initial tangential force distribution (when no external tangential load has been yet applied). In the model, such forces are assumed to be zero all along the contact. However, in the experiments, both the slider and the base undergo different expansion rates during application of the normal loading. As discussed in~\cite{Rubinstein-Cohen-Fineberg-JPhysD-2009}, the associated slip at the interface is impeded by friction, thus yielding a significant tangential force distribution at the interface. Such distributions have been measured to be antisymmetric~\cite{BenDavid-Cohen-Fineberg-Science-2010}, in agreement with basic contact mechanics calculations~\cite{Johnson-CUP-1985}, thus ensuring $F_T(0)=\sum_{n=1}^N \tau_n (0) = 0$. This effect is again a bulk effect which is quantitatively reproduced in 2D models~\cite{Tromborg-Scheibert-Amundsen-Thogersen-MaltheSorenssen-PhysRevLett-2011}. Here, in 1D, we will only study the qualitative influence of an initial tangential force distribution on the length of precursors. We consider the simple linear distributions shown in Fig.~\ref{fig:initialShearForce}a.

Implementation of an initial tangential force requires an initial relative displacement of the blocks. The initial tangential forces are given by
\begin{equation}
\tau_n(0) = \left\{ \begin{array}{ll}
k(u_2(0) - u_1(0)) +  F_T(0), & n = 1 \\
k(u_{n+1}(0) - 2u_n(0) + u_{n-1}(0)), & 2 \leq n \leq N-1 \\
k(u_{N-1}(0) - u_N(0)), & n = N,
\end{array} \right.
\end{equation}
and by choosing $F_T(0) = 0$ and $u_1(0) = 0$ the above equation can be rewritten to
\begin{equation}
u_n(0) = \left\{ \begin{array}{ll}
0, & n = 1 \\
u_1 + \tau_n/k, & n = 2 \\
2u_{n-1} - u_{n-2} + \tau_{n-1}/k, & n = 3,4,\dotsc,N ,
\end{array} \right.
\label{eq:initialPositions}
\end{equation}
thus enabling calculation of the initial positions of all blocks given $\tau_n$. The initial attachment position of the interfacial springs is then calculated using Eq.~\eqref{eq:u_n^stick}, which ensures that the total force on each block is zero at $t = 0$.

The length of precursors corresponding to the three initial tangential force profiles shown in Fig.~\ref{fig:initialShearForce}a is shown in Fig.~\ref{fig:initialShearForce}b, using $\theta = 0$. With respect to a zero initial force distribution, the stronger the asymmetry the lower the $L_p$ \textit{vs.} $F_T$ curve, \textit{i.e.} the shorter the precursors for the same tangential force. Moreover, the slope of the curves is significantly increased at large forces. Both effects lead to a significantly improved qualitative agreement with the experimental results by \changed{both Maegawa \textit{et al.} and Rubinstein \textit{et al.}~\cite{Rubinstein-Cohen-Fineberg-PhysRevLett-2007}}. This shows that the initial force distribution, which arises naturally in 2D models~\cite{Tromborg-Scheibert-Amundsen-Thogersen-MaltheSorenssen-PhysRevLett-2011}, is a crucial parameter for the kinematics of precursors to sliding. Note that the number of precursors, which is closely related to the choice of $l_0$ (as mentioned in the previous subsection) is only weakly affected by the introduction of an initial tangential force distribution.


\section{Analytical prediction of precursor lengths}
\label{sec:AnalyticPredictionOfPrecursorLengths}

In order to complete this study we derive an analytic prediction for the precursor length as a function of the tangential force at event arrest, $L_p$ \textit{vs.} $F_T$. We will first look at the simpler case of the model by Maegawa \textit{et al.}, and then extend the prediction to our improved model using both an interfacial stiffness and an initial tangential force profile.

\subsection{Prediction in the model by Maegawa \textit{et al.}}

Assume that a precursor has reached block $n_p$ and has the length $L_p = (n_p/N)L$. We want to calculate the tangential force $F_T$ at the time of arrest of this event. At that time, all blocks are stuck, so that
\begin{equation}
F_T = \sum_{n=1}^N \tau_n.
\label{eq:F_T}
\end{equation}
This means that, given the tangential force distribution at event arrest, the corresponding tangential force is found using Eq.~\eqref{eq:F_T}. 

According to Figs.~\ref{fig:shearForceProfileNoVistosityAmontonsCoulombN=100} and~\ref{fig:shearForceProfileWithVistosityAmontonsCoulombN=100} the tangential force is observed to be approximately equal to the kinetic friction level from block $1$ to $n_p$, and $0$ elsewhere. Using this assumption, Eq.~\eqref{eq:F_T} yields
\begin{equation}
F_T = \mu_k \sum_{n=1}^{n_p} p_n =
\mu_k \frac{F_N}{N} \sum_{n=1}^{n_p} 1 - \frac{2n - N - 1}{N - 1} \theta,
\end{equation}
where Eq.~\eqref{eq:p_n} has been inserted for $p_n$. If $N \gg 1$, the sum can be approximated by an integral and $n$ replaced by $x = nL/N$, which yields
\changed{
\begin{equation}
F_T \approx \mu_k \frac{F_N}{N} \frac{N}{L} \int_0^{L_p} \left[ 1 - \frac{2(x N/L) - N - 1}{N - 1} \theta \right] \, \md x,
\end{equation}
}
and approximating $N \pm 1 \approx N$ yields
\changed{
\begin{align}
F_T &\approx \mu_k \frac{F_N}{L} \int_0^{L_p} \left[ 1 - \left( 2(x/L) - 1 \right) \theta \right] \, \md x \\
F_T &\approx \mu_k F_N \frac{L_p}{L} \left[ 1 + \theta \left( 1 - \frac{L_p}{L} \right) \right] .
\label{eq:AC:F_T(L_p)}
\end{align}
}
As seen in Fig.~\ref{fig:precursorLengthsNoInitialShearForceAmontonsCoulomb}, this prediction is in very good agreement with our simulation results. The deviation between the actual precursors and the analytical curve is the result of a slightly incorrect assumed tangential force profile. Note that a similar good prediction scheme, numerical rather than analytical, was previously developed in the 2D study by Tr\o mborg \textit{et al.}~\cite{Tromborg-Scheibert-Amundsen-Thogersen-MaltheSorenssen-PhysRevLett-2011}.

According to Eq.~\eqref{eq:AC:F_T(L_p)} the global static friction coefficient $\mu_S$ ($\mu_S \approx \frac{F_T}{F_N}$ is easily evaluated for $L_p=L$ from Eq.~\eqref{eq:AC:F_T(L_p)}) is independent of $\theta$ and almost equal to the local kinetic friction coefficient $\mu_k$. This is in agreement with the model result in Fig.~\ref{fig:precursorLengthsNoInitialShearForceAmontonsCoulomb} and with the results of the 2D model by Tr\o mborg \textit{et al.}~\cite{Tromborg-Scheibert-Amundsen-Thogersen-MaltheSorenssen-PhysRevLett-2011}. 

\subsection{Prediction in our improved model}

Prediction of the precursor length in our improved model including both an interfacial stiffness and an initial tangential force profile follows the same line as that for the model by Maegawa \textit{et al.}. However, blocks may now move even though no event is occurring. Despite this, it is expected that block accelerations are small when all track springs are attached, which leads to the approximate validity of Eq.~\eqref{eq:F_T}.

\begin{figure}
\centering
\includegraphics{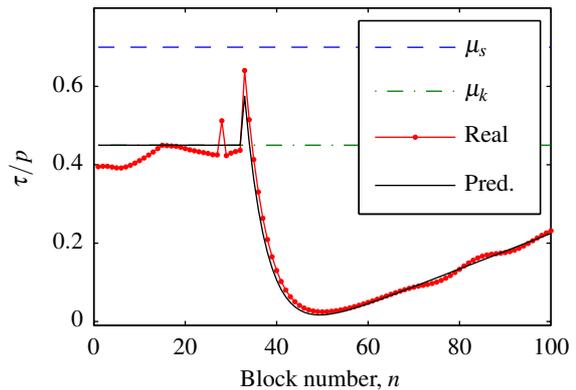}
\caption{The tangential force profile at the arrest of the $14$th precursor plotted in Fig.~\ref{fig:initialShearForce} (red dots) and the predicted profile used in the analytical prediction of precursor lengths (black drawn line) using $N = 100$, $\theta = 0$, $\eta = \sqrt{0.1}\sqrt{km}$, $l_0 = 5$~mm and $\beta = 0.225$ in Eq.~\eqref{eq:tau^0(x)}.}
\label{fig:shearForceProfileWithAnalytical}
\end{figure}

An approximate tangential force profile at the time of arrest of an event (Fig.~\ref{fig:shearForceProfileWithAnalytical}) has to be found for a given $L_p$. Again, blocks in $[0, L_p]$ are assumed to have a tangential force equal to the kinetic friction level. Blocks in the interval $[L_p, L]$, however, now needs to be taken into account for two reasons: both the initial tangential force profile and the springs to the track lead to a non-zero tangential force for $x > L_p$. The form of this profile in a static situation has been calculated in App.~\ref{sec:ScalingOfTheModelWithTheSpringToTrackStaticFrictionLaw}, and is given in Eq.~\eqref{eq:shearForceProfile}, where $\tau^0(x)$ now is the initial tangential force profile. However, it has to be modified to take into account that the loaded block is not located at $x = 0$, but at $x = L_p$, and that this block does not have to be loaded up to the static friction level, but may take some other value, say $\alpha p_{n_p}$, with $\alpha$ a coefficient to be defined. Our assumed tangential force profile at the arrest of a precursor of length $L_p$ is therefore given by
\begin{equation}
\tau(x) = \left\{ \begin{array}{ll}
\mu_k p(x), & x \in \left[0, L_p\right] \\
\left(\alpha p(L_p) - \tau^0 (x) \right) e^{-\frac{x - L_p}{l_0}} + \tau^0(x), & x \in \left[ L_p, L \right]
\end{array} \right.
\label{eq:ours:F_T(L_p)}
\end{equation}
We have considered that all values of the amplitude of the peak at $x=L_p$ have the same probability to occur between $\mu_k p$ and $\mu_s p$, so that we have given $\alpha$ its average value $\alpha = (\mu_s + \mu_k)/2$ in the predictions seen in Fig.~\ref{fig:initialShearForce}b. Figure~\ref{fig:shearForceProfileWithAnalytical} shows the assumed tangential force profile and the actual tangential force profile at the arrest of an event, and the agreement is seen to be satisfactory.

From Eqs.~\eqref{eq:ours:F_T(L_p)}, one can calculate $F_T(L_p)$ in the same way as was done for the model by Maegawa \textit{et al.} This calculation is provided in Appendix~\ref{sec:ours:precursorPrediction}. As seen in Fig.~\ref{fig:initialShearForce}b, the prediction scheme works well. Deviations between the actual precursors and the analytical curve have two contributions: incorrectly assumed tangential force profile and inertial effects, where the former gives the largest contribution. Again, the global static friction coefficient $\mu_S$ is seen to be approximately equal to the local kinetic friction coefficient $\mu_k$. The prediction curves appear to bend slightly backwards at $F_T/F_N \sim 0.45$. In our continuous prediction scheme, this corresponds to micro-slip fronts that are so close to the leading edge that all blocks at $x > L_p$ have a tangential force above the kinetic friction level, causing $F_T$ to get smaller as the front moves further. In reality, however, no precursor ever arrests in such a state but propagates all the way to the trailing edge.


\section{Conclusion}

Recent experimental results about the transition from static to kinetic friction in line contacts have triggered the study of the deterministic dynamics of 1D spring-block friction models in which driving is applied at one extremity of the chain of blocks. In this Letter, we have improved the simplest of such models~\cite{Maegawa-Suzuki-Nakano-TribolLett-2010} in order to solve its intrinsic unphysical resolution-dependence and to ameliorate its qualitative agreement with experimental results on the kinematics of micro-slip fronts along the contact. In particular, the introduction of a tangential stiffness of the interface, by introducing a new length scale in the model, practically suppresses its resolution dependence and allows for reproduction of realistic numbers of precursory micro-slip fronts. The additional introduction of an initial tangential force distribution at the interface significantly improves the agreement with the evolution of the precursor length with the external tangential load obtained in experiments. Our improved model is intented to serve as a framework for more complex friction models when robust comparisons with experiments are desired.

We focused on 1D models because they are simple enough to enable deep insights into the qualitative effects of the model's parameters. However, it is known that 2D models~\cite{Tromborg-Scheibert-Amundsen-Thogersen-MaltheSorenssen-PhysRevLett-2011} are required to provide quantitative agreement with experiments. In this respect, the improvements brought to the 1D model are effective ways to account for intrinsically 2D effects: First, the length scale introduced through the interfacial stiffness enables coupling between remote points along the interface, analogous to the coupling through the slider's bulk; Second, the initial tangential force distribution accounts for the shear stress arising form the differential Poisson expansion of two bodies pressed together.

In analogy with 2D results we developed, based on the well defined force distribution left by an arrested precursor, an efficient analytical prediction for the precursors' length as a function of the external tangential load applied. We also find that, like in 2D, the macroscopic static friction coefficient of a side-driven contact is approximately equal to its microscopic kinematic friction coefficient.


\begin{acknowledgements}
We thank J. L. Vinningland for discussions. We acknowledge funding from the European Union (Marie Curie Grant No. PIEF-GA-2009-237089). This article was supported by a Center of Excellence grant to PGP from the Norwegian Research Council.
\end{acknowledgements}


\appendix

\section{Relative viscous damping in a linear chain of blocks}
\label{sec:RelativeViscousDampingInALinearChainOfBlocks}

If friction forces are ignored, the equation of motion for an infinite chain of blocks connected by springs is given by
\begin{equation}
m\ddot{u}_n = k(u_{n + 1} - 2u_n + u_{n-1}) + \eta(\dot{u}_{n+1} - 2\dot{u}_n + \dot{u}_{n-1}) .
\label{eq:1DinfiniteChainDiffEq}
\end{equation}
We then assume a solution of the form
\begin{equation}
u_n(t) = e^{\zeta_\kappa t} e^{i\kappa n a},
\label{eq:1DinfiniteChainSolution}
\end{equation}
where $\zeta_\kappa \in \mathbb{C}$ and $\kappa \in \mathbb{R}$. Inserting Eq.~\eqref{eq:1DinfiniteChainSolution} into Eq.~\eqref{eq:1DinfiniteChainDiffEq} yields the relation
\begin{equation}
m \zeta_\kappa^2 =
k \left( e^{i\kappa a} - 2 + e^{-i\kappa a} \right) + \eta \zeta_\kappa \left( e^{i\kappa a} - 2 + e^{-i\kappa a} \right) ,
\label{eq:infiniteChainEquation1}
\end{equation}
which can be simplified to
\begin{equation}
m \zeta_\kappa^2
+ 4 \eta \sin^2 \left( \frac{\kappa a}{2} \right) \zeta_\kappa + 4k \sin^2 \left( \frac{\kappa a}{2} \right) = 0 ,
\label{eq:infiniteChainEquation2}
\end{equation}
since
\begin{equation} e^{i\kappa a} - 2 + e^{-i\kappa a} = -4\sin^2 \left( \frac{\kappa a}{2} \right) .
\end{equation}
The complex parameter $\zeta_\kappa$ is then given by
\begin{equation}
\zeta_\kappa = \frac{-4 \eta \sin^2 \left( \frac{\kappa a}{2} \right) \pm \sqrt{16 \eta^2\sin^4 \left( \frac{\kappa a}{2} \right) -16 km \sin^2 \left( \frac{\kappa a}{2} \right)}}{2m}
\end{equation}
The system is critically damped when Eq.~\eqref{eq:infiniteChainEquation2} only has one solution for $\zeta_\kappa$, which occurs when the square root is zero:
\begin{equation}
\eta^2 \sin^2 \left( \frac{\kappa a}{2} \right) = km \quad \Rightarrow \quad
\eta = \frac{\sqrt{km}}{\left| \sin \left( \frac{\kappa a}{2} \right) \right|} .
\label{eq:dampingEq3}
\end{equation}
The oscillations that are to be reduced have a wavelength $\lambda = 2a$, \textit{i.e.} a wave number $\kappa = 2\pi/\lambda = \pi/a$. Inserting this into Eq.~\eqref{eq:dampingEq3} leads to
\begin{equation}
\eta_c = \sqrt{k m},
\end{equation}
which is the value of the damping coefficient $\eta$ for which waves of wavelength $\lambda = 2a$ are critically damped. Since the absolute value of $\sin$ is always smaller than one, choosing $\eta = \sqrt{km}$ will cause all other waves to be under-damped.


\section{Tangential force profiles and characteristic length with a tangential stiffness of the interface}
\label{sec:ScalingOfTheModelWithTheSpringToTrackStaticFrictionLaw}

An analytical expression for the characteristic length $l_0$ can be found. In order to do so, the following assumptions are made: $N \gg 1$, $l_0/L \ll 1$ and
slow loading compared to the internal dynamics of the system, which enables a static analysis. The system is first placed in a static state with an initial shear force profile given by $\tau_n^0$, and then loaded slowly from the left. The equilibrium of all non-edge blocks writes
\begin{equation}
k\left( u_{n+1} - 2u_n + u_{n-1} \right) - k_t \left( u_n - u_n^\text{stick} \right) = 0.
\label{eq:staticEquation1}
\end{equation}
We introduce a new variable $u_n'$ defined by
\begin{equation}
u_n = u_n' + u_n^0 ,
\label{eq:newBlockPositionVariable}
\end{equation}
where $u_n^0$ is the initial position of block $n$. Inserting Eq.~\eqref{eq:newBlockPositionVariable} into Eq.~\eqref{eq:staticEquation1} yields
\begin{equation}
k\left( u_{n+1}' - 2u_n' + u_{n-1}' \right) -
k_t u_n' + \tau_n^0 - k_t \left( u_n^0 - u_n^\text{stick} \right) = 0,
\label{eq:staticEquation2}
\end{equation}
where
\begin{equation}
\tau_n^0 = k\left( u_{n+1}^0 - 2u_n^0 + u_{n-1}^0 \right) .
\end{equation}
The two terms $\tau_n^0$ and $-k_t \left( u_n^0 - u_n^\text{stick} \right)$ cancel in Eq.~\eqref{eq:staticEquation2} since the initial state is static, and thus
\begin{equation}
k\left( u_{n+1}' - 2u_n' + u_{n-1}' \right) -
k_t u_n' = 0.
\label{eq:staticEquation3}
\end{equation}
The above equation can be rewritten to
\begin{equation}
k a^2 \frac{u_{n+1}' - 2u_n' + u_{n-1}'}{a^2} -  k_t u_n' = 0,
\label{eq:staticEquation4}
\end{equation}
where $a = L/(N_x - 1)$ is the lattice spacing. Since $N \gg 1$, the first term in Eq.~\eqref{eq:staticEquation4} can be replaced by the second spatial derivative, and replacing $u_n'$ with $u'(na) = u'(x)$ yields
\begin{equation}
k a^2 \frac{\partial^2 u'(x)}{\partial x^2} - k_t u'(x) = 0 ,
\end{equation}
which has the general solution
\begin{equation}
u'(x) = A e^{x/l_0} + Be^{-x/l_0}, \quad
l_0 = \sqrt{\frac{k}{k_t}}a .
\label{eq:staticSolution1}
\end{equation}
The shear force is given by
\begin{align}
\tau_n &= k \left( u_{n+1} - 2u_n + u_{n-1} \right) \\
&= k \left( u_{n+1}' - 2u_n' + u_{n-1}' \right) + \tau_n^0 .
\end{align}
By replacing again finite differences with second order derivatives,
\begin{equation}
\tau(x) = ka^2 \frac{\partial^2 u'(x)}{\partial x^2} + \tau^0(x),
\end{equation}
and the general expression for the shear force profile can be found by using Eq.~\eqref{eq:staticSolution1}, which yields
\begin{equation}
\tau(x) = \frac{ka^2 l_0^2}{L^2} \left( A e^{x/l_0} + Be^{-x/l_0} \right) + \tau^0(x) .
\end{equation}
The system is loaded from the left, and at the beginning of an event the shear force on block $1$ is equal to the static friction threshold $\mu_s p_1$. Provided \changed{$l_0/L \ll 1$}, the trailing edge will not be affected by the loading. The latter of these two boundary conditions yields
\begin{align}
\tau(L) &= \frac{ka^2 l_0^2}{L^2} \left( A e^{L/l_0} + Be^{-L/l_0} \right) + \tau^0(L) \\
&\approx \frac{ka^2 l_0^2}{L^2} \left( A e^{L/l_0} \right) + \tau^0(L) \\
&= \tau^0 (L), 
\end{align}
i.e. $A = 0$. The first boundary condition yields
\begin{equation}
\tau(0) = \frac{ka^2 l_0^2}{L^2} B + \tau^0(L) = \mu_s p_1,
\end{equation}
and the shear force is therefore given by
\begin{equation}
\tau(x) = \left( \mu_s p_1 - \tau^0 (x) \right)e^{-x/l_0} + \tau^0(x) .
\label{eq:shearForceProfile}
\end{equation}

The characteristic length $l_0$ is given by Eq.~\eqref{eq:staticSolution1}, and inserting for $k$ given by Eq.~\eqref{eq:k} and $a$ yields
\begin{align}
l_0 &= \sqrt{\frac{k}{k_t}}a = \sqrt{\frac{E L S}{(N - 1) k_t}},
\end{align}
and hence Eq.~\eqref{eq:l0N} for $N>>1$.

\changed{Note that in a 3D situation, the exponential decay of the tangential stress with $x$ would be replaced by a power law~\cite{Landau-Lifshitz-Pergamon-1986, Braun-Barel-Urbakh-PhysRevLett-2009}.}


\section{Derivation of the prediction of precursor lengths in our improved model}
\label{sec:ours:precursorPrediction}

We start with Eq.~\eqref{eq:F_T} and use the assumed shear force profile in Eq.~\eqref{eq:ours:F_T(L_p)}, shown in Fig.~\ref{fig:shearForceProfileWithAnalytical}. Again we go to the limit $N \to \infty$, resulting in the substitution
\begin{equation}
\sum_{n=1}^N \tau_n \to \frac{N}{L} \int_0^L \tau(x) \, \md x , \qquad
n \to xN/L .
\end{equation}
The tangential force after a precursor of length $L_p$ is then given by
\begin{align}
F_T &= \frac{N}{L} \left[ \int_0^{L_p} \tau(x) \, \md x + \int_{L_p}^{L} \tau(x) \, \md x \right] \\
&= \frac{N}{L} \left[ \int_0^{L_p} \mu_k p(x) \, \md x +
\int_{L_p}^{L} \left(\alpha p(L_p) - \tau^0 (x) \right) e^{-\frac{x - L_p}{l_0}} + \tau^0(x) \, \md x \right] .
\label{eq:F_T1}
\end{align}
We limit ourselves to predicting the precursors in Fig.~\ref{fig:initialShearForce}b, \textit{i.e.} using a tangential interfacial stiffness and a linear initial shear forces as depicted in Fig.~\ref{fig:initialShearForce}a, but with $\theta = 0$. The normal and initial shear force are then given by
\begin{align}
p(x) &= p = F_N/N = \text{constant}, \label{eq:p(x)} \\
\tau^0 (x) &= \beta p \frac{2(x - L/2)}{L},
\label{eq:tau^0(x)}
\end{align}
where the parameter $\beta$ determines the slope in the initial shear force profile. Inserting Eqs.~\eqref{eq:p(x)} and~\eqref{eq:tau^0(x)} into Eq.~\eqref{eq:F_T1} yields
\begin{multline}
F_T = \frac{N}{L} \left[ \int_0^{L_p} \mu_k p \, \md x \, + \right. \\
\left. \int_{L_p}^{L} \left(\alpha p - \beta p \frac{2(x - L/2)}{L} \right) e^{-\frac{x - L_p}{l_0}} + \beta p \frac{2(x - L/2)}{L} \, \md x \right] .
\end{multline}
The above integrals can be calculated easily, and the result is the tangential load $F_T$ as a function of the precursor length $L_p$:
\begin{multline}
F_T(L_p) = F_N \left[ \mu_k \frac{L_p}{L} + 2\beta \frac{l_0^2}{L^2} \left(e^{-\frac{L-L_p}{l_0}} - 1 \right) + \beta \frac{\left(L-L_p\right) L_p}{L^2} + \right. \\
\left. \frac{l_0}{L} \left( \beta \left(1+e^{-\frac{L-L_p}{l_0}} - 2\frac{L_p}{L} \right) +
\alpha  \left(1 - e^{-\frac{L-L_p}{l_0}}\right) \right) \right] .
\label{eq:F_T(L_p)_springToTrack}
\end{multline}
We observe that again $F_T(L) = \mu_k F_N$.

\end{document}